\documentclass[sigconf,screen]{acmart}

\usepackage{multirow}
\usepackage{balance}
\usepackage{diagbox}
\usepackage{color}
\usepackage{censor}
\usepackage{url}
\usepackage{booktabs}
\newcommand{\toolname}{CIRCLE}

\usepackage{subfigure}
\usepackage{tablefootnote}
\usepackage{colortbl}

\AtBeginDocument{%
  \providecommand\BibTeX{{%
    \normalfont B\kern-0.5em{\scshape i\kern-0.25em b}\kern-0.8em\TeX}}}

\setcopyright{acmcopyright}
\acmPrice{15.00}
\acmDOI{10.1145/3533767.3534219}
\acmYear{2022}
\copyrightyear{2022}
\acmSubmissionID{issta22main-p107-p}
\acmISBN{978-1-4503-9379-9/22/07}
\acmConference[ISSTA '22]{Proceedings of the 31st ACM SIGSOFT International Symposium on Software Testing and Analysis}{July 18--22, 2022}{Virtual, South Korea}
\acmBooktitle{Proceedings of the 31st ACM SIGSOFT International Symposium on Software Testing and Analysis (ISSTA '22), July 18--22, 2022, Virtual, South Korea}

\begin{document}

\title{CIRCLE: Continual Repair across Programming Languages}

\author{Wei Yuan}
\authornote{Both authors contributed equally to this research.}
\email{w.yuan@uq.edu.au}
\affiliation{%
  \institution{School of Information Technology and Electrical Engineering}
  \city{The University of Queensland}
  \country{Australia}
}

\author{Quanjun Zhang}
\authornotemark[1]
\email{quanjun.zhang@smail.nju.edu.cn}
\affiliation{%
  \institution{State Key Laboratory for Novel Software Technology}
  \city{Nanjing University}
  \country{China}
}

\author{Tieke He}
\authornote{Corresponding authors.}
\email{hetieke@gmail.com}
\affiliation{%
  \institution{State Key Laboratory for Novel Software Technology}
  \city{Nanjing University}
  \country{China}
}

\author{Chunrong Fang}
\authornotemark[2]
\email{fangchunrong@nju.edu.cn}
\affiliation{%
  \institution{State Key Laboratory for Novel Software Technology}
  \city{Nanjing University}
  \country{China}
}

\author{Nguyen Quoc Viet Hung}
\email{henry.nguyen@griffith.edu.au}
\affiliation{%
  \institution{Institute for Integrated and Intelligent Systems}
  \city{Griffith University}
  \country{Australia}
}

\author{Xiaodong Hao}
\email{mf21320054@smail.nju.edu.cn}
\affiliation{%
  \institution{State Key Laboratory for Novel Software Technology}
  \city{Nanjing University}
  \country{China}
}

\author{Hongzhi Yin}
\email{h.yin1@uq.edu.au}
\affiliation{%
  \institution{School of Information Technology and Electrical Engineering}
  \city{The University of Queensland}
  \country{Australia}
}

\renewcommand{\shortauthors}{Yuan and Zhang, et al.}

\begin{abstract}
  Automatic Program Repair (APR) aims at fixing buggy source code with less manual debugging efforts, which plays a vital role in improving software reliability and development productivity.
  Recent APR works have achieved remarkable progress via applying deep learning (DL), particularly neural machine translation (NMT) techniques.
  However, we observe that existing DL-based APR models suffer from at least two severe drawbacks:
  (1) Most of them can only generate patches for a single programming language, as a result, to repair multiple languages, we have to build and train many repairing models.
  (2) Most of them are developed offline. Therefore, they won't function when there are new-coming requirements.
 
  To address the above problems, a T5-based APR framework equipped with continual learning ability across multiple programming languages is proposed, namely \emph{C}ont\emph{I}nual \emph{R}epair a\emph{C}ross Programming \emph{L}anguag\emph{E}s (\emph{CIRCLE}).
  Specifically,
  (1) CIRCLE utilizes a prompting function to narrow the gap between natural language processing (NLP) pre-trained tasks and APR.
  (2) CIRCLE adopts a difficulty-based rehearsal strategy to achieve lifelong learning for APR without access to the full historical data. 
  (3) An elastic regularization method is employed to strengthen CIRCLE's continual learning ability further, preventing it from catastrophic forgetting.
  (4) CIRCLE applies a simple but effective re-repairing method to revise generated errors caused by crossing multiple programming languages.
  
  We train CIRCLE for four languages (i.e., C, JAVA, JavaScript, and Python) and evaluate it on five commonly used benchmarks. The experimental results demonstrate that CIRCLE not only effectively and efficiently repairs multiple programming languages in continual learning settings, but also achieves state-of-the-art performance (e.g., fixes 64 Defects4J bugs) with a single repair model.
\end{abstract}




\maketitle

\section{Introduction}\label{sec_intro}
Automatic Program Repair (APR) is critical for software developers since manually detecting and fixing bugs is a labor-intensive and time-consuming task~\cite{weiss2007long}.
With the recent advances in deep learning (DL), a lot of APR approaches have been proposed to use neural network techniques to learn the bug-fixing patterns from accessible code repositories~\cite{gazzola2017automatic,bader2019getafix,ye2021comprehensive}.
Assisted by deep learning's powerful ability to learn hidden and intricate relationships from massive data, DL-based APRs achieve remarkable performance \cite{Zhu2021recoder,jiang2021cure}.

Generally, the DL-based APRs is composed by two parts~\cite{cho2014learning}:
an encoder that extracts the meaning of buggy code with necessary context and converts it into fix-length vectors; 
and a decoder that generates the correct statements from the encoder's output~\cite{long2016automatic,jiang2018shaping,li2020dlfix}.
These encoder-decoder models usually treat the program repair task as a translation from buggy code to fixed code.
For example, Tufano et al.~\cite{tufano2019empirical} adopt a classical neural machine translation (NMT) model to generate the fix patches.
CoCoNut~\cite{lutellier2020coconut} utilizes two separate encoders to encode the buggy lines and surrounding context.
Further, CURE~\cite{jiang2021cure} employs GPT to provide contextual embeddings for CoCoNut.
Zhu et al.~\cite{Zhu2021recoder} design a syntax-guided edit decoder to generate patches in a refined way.

Despite the recent achievements, existing DL-based APR techniques have at least two limitations.
First, \emph{most of them can only fix bugs for a single language.} 
If we want to repair codes for many languages, we have to train and store a corresponding number of models.
In addition, these models are trained independently, which limits them to transfer the latent knowledge learned from other languages.
We argue that such ``single-language learning setting'' is artificial because the underlying code understanding and bug identifying abilities may share a large common across languages.
For example, a developer who is proficient in certain language, when (s)he reads code with unfamiliar language, (s)he can still understand the general ideas and intuitively point out the potential problem.
Recent research also indicates that multilingual training data could improve performance on several code-related tasks (e.g., code summarization and method name prediction)~\cite{ahmed2021multilingual, Zhu2022}.
Therefore, learning multiple languages fixing may be more efficient and effective than separately learn different languages.
Besides, the scalability of cross-lingual repairing model will also be better than these traditional APR models, since the prior one can handle various languages with just one model.
And the scalability problem will probably be severer with the growing of neural network models' size.

Another shortcoming for current APRs is that \emph{they are developed in an offline manner}, so they cannot continually improve their bug-fixing ability.
This shortcoming decreases the value of DL-based APRs in real-world scenarios where new task requirements increase constantly.
The ``task requirements increase constantly'' refers to that only a part of tasks are targeted to be solved at the first time and new task requirements are proposed afterwards.
This case happens when tackling all tasks at once is very complex and time-consuming or when the task requirements cannot be fully obtained initially.
For example, when companies decide to provide APR service, they tend to provide service for the most in-demand programming language in the first place, since creating a high-quality and enterprise-level APR model is costly.
Later, as the business grew, they would like to enrich their APR service for other languages.
In addition, collecting bug-fixing corpus is also an adaptive and continual process, even though all task requirements are covered initially, the model still needs to expand their knowledge on new corpus.
Conventional APR models tend to overwrite the knowledge learned from previous tasks when learning new tasks.
As a result, every time the task requirement increases, APR models have to be retrained on all corpus, which is time-consuming.

To mitigate the above issues, a new DL-based APR model that can process multi-type programming languages and continually learn defects fixing is desirable.
However, there are two main challenges in implementing such APR models.
First, repairing defects cross languages is more difficult than for a single language~\cite{van2019towards}.
Therefore, it is essential to efficiently and effectively exploit the power of deep learning~\cite{liu2021pre}, especially the large pre-trained neural network models, which are commonly used in NLP~\cite{qiu2020pre,wolf2020transformers,otter2020survey}.
Second, models are prone to forget the knowledge obtained from previous tasks when they are learning on new corpus or new tasks, i.e., they are struggling with catastrophic forgetting\footnote{Catastrophic forgetting means that neural network model tends to completely forget learned knowledge when learning new information.}~\cite{mccloskey1989catastrophic,french1999catastrophic}.
How to prevent this forgetting is non-trivial.

In this paper, we propose \emph{CIRCLE} (short for \underline{C}ont\underline{I}nual \underline{R}epair a\underline{C}ross Programming \underline{L}anguag\underline{E}s), which is able to continually learn bug fixing across multiple programming languages.
To be specific, CIRCLE incorporates a prompt template, a T5-based APR model, and a re-repairing mechanism to address the first limitation and challenge mentioned above (i.e. repairing across languages and effectively utilizing pre-trained models).
Concretely, T5~\cite{raffel2019exploring} is a widely used NLP pre-trained model that exhibits a formidable capacity for handling multiple tasks~\cite{mastropaolo2021studying,ahmad2021unified}.
The prompt template converts bug-fixing inputs into fill-in-the-blank form, closing the gap between T5's pre-trained task and program repair.
This prompt template helps model better exploit the knowledge learned from pre-trained tasks~\cite{shin2020autoprompt,schick2020exploiting,liu2021gpt,lester2021power}.
The re-repairing mechanism is designed to eliminate incorrectly generated patches caused by crossing languages.

To tackle the second issue and challenge (i.e., continually learning bug-fixing without catastrophic forgetting), CIRCLE applies a rehearsal method and an elastic regularization.
The rehearsal method stores a small set of data from past datasets to simulate the historical data distribution and replays them in the later learning period.
The main challenge is how to select the set of ``representative data''.
CIRCLE proposes a novel data selection scheme that collects representative examples from historical data for bug repairing based on the difficulty.
However, since the size of selected samples is desired to be small enough to reduce the resource costs, solely using the difficulty-based example replay method cannot adequately avoid the forgetting problem.
Therefore, CIRCLE employs a parameter updating regularization approach based on Elastic Weight Consolidation (EWC)~\cite{kirkpatrick2017overcoming}.
Moreover, CIRCLE calculates the Fisher Matrix~\cite{thompson2019overcoming} on the chosen examples rather than on the whole historical data to approximate EWC values so that it does not need to keep the whole historical data.
The EWC imposes restrictions on parameters that play a vital role in previous task learning, forcing model to learn current and future tasks via adapting other parameters.

Extensive experiments are conducted across $4$ popular programming languages (C, JAVA, JavaScript, and Python) on $5$ benchmarks to test the effectiveness of our CIRCLE.
The experimental results demonstrate that a \emph{single} CIRCLE model can continually learn program repair crossing multiple languages settings without severely forgetting previous knowledge.
Furthermore, CIRCLE outperforms the conventional state-of-the-art DL-based APRs which are dedicated trained for certain language.
At last, each component's importance is studied in the ablation study in detail.
The code, experimental results and processed data is available\footnote{\url{https://github.com/2022CIRCLE/CIRCLE}.}.

To sum up, the main contributions of this paper are three-fold:
\begin{itemize}
  \item We propose CIRCLE, a novel program repair framework that can continually learn bug-fixing across multiple languages. To the best of our knowledge, we are the first to explore multi-language program repair in continual learning scenarios.
  \item A prompt-based template is developed to convert program repair into ``fill-in-the-blank'' task, allowing the pre-trained model, T5, to perform the repair task effectively. A simple but effective re-repairing mechanism is designed to revise generation errors about crossing languages. In addition, a novel difficulty-based example replay and an EWC-based regularization method are proposed to mitigate catastrophic forgetting in continual learning settings.
  \item Extensive experiments are conducted with $4$ commonly used programming languages and evaluated on $5$ bug benchmarks to demonstrate that CIRCLE can continually learn bug repair crossing languages and outperform previous neural APR models. Ablation studies and further analyses are presented to discuss CIRCLE's performance.
\end{itemize}  

The remainder of this paper is organized as follows. 
Section 2 provides the basic background related to our work. 
Section 3 introduces the details of our CIRCLE.
Section 4 describes the datasets and metrics adopted in this paper, followed by experimental results and discussions.
Section 5 presents the related work on program repair and lifelong learning.
Section 6 concludes our work.

\section{Background}
\subsection{DL-based APR}
DL-based APRs have achieved state-of-the-art performance on program repair task~\cite{chen2019sequencer,lutellier2020coconut,tufano2019learning,chakraborty2020codit,ye2021neural}.
Most of them treat repairing as a neural machine translation task and optimize an encoder-decoder model on a set of bug-fix pairs to learn latent patterns based on supervised learning.
The inputs and neural model architectures of DL-based APRs are various.
For example, CoCoNut~\cite{lutellier2020coconut} separately encodes context and buggy codes by CNN networks.
SequenceR~\cite{chen2019sequencer} abstracts the buggy context and takes it as input together with buggy lines.
Recently, pre-trained models are also used in DL-based APRs. 
CURE~\cite{jiang2021cure} employs GPT as token embedding layer.
Mashhadi et al.~\cite{mashhadi2021applying} utilize CodeBERT to fix Java simple bugs.

However, to the best of our knowledge, repairing multiple programming languages' defects via a single model is still underexplored.
In light of this, we propose to employ the recent pre-trained model T5 as the skeleton and build a repair model that can fix bugs across languages.

\subsection{Continual Learning}~\label{sec:cl}
Continual Learning (also referred to as Lifelong Learning) is a type of machine learning paradigm, which aims to continually learn new tasks while not severely forget previously gained knowledge~\cite{parisi2019continual}.
Formally, let $f_{t}$ denote the model trained in task $t$, the new dataset $D_{t+1}$ for task $t+1$ is used to update the model $f_{t}$.
Continual Learning attempts to ensure that after updating, model $f_{t+1}$ can have good performance in all seen tasks $1:t+1$.

\begin{figure}[!t]
  \centering
  \includegraphics[width=3.3in]{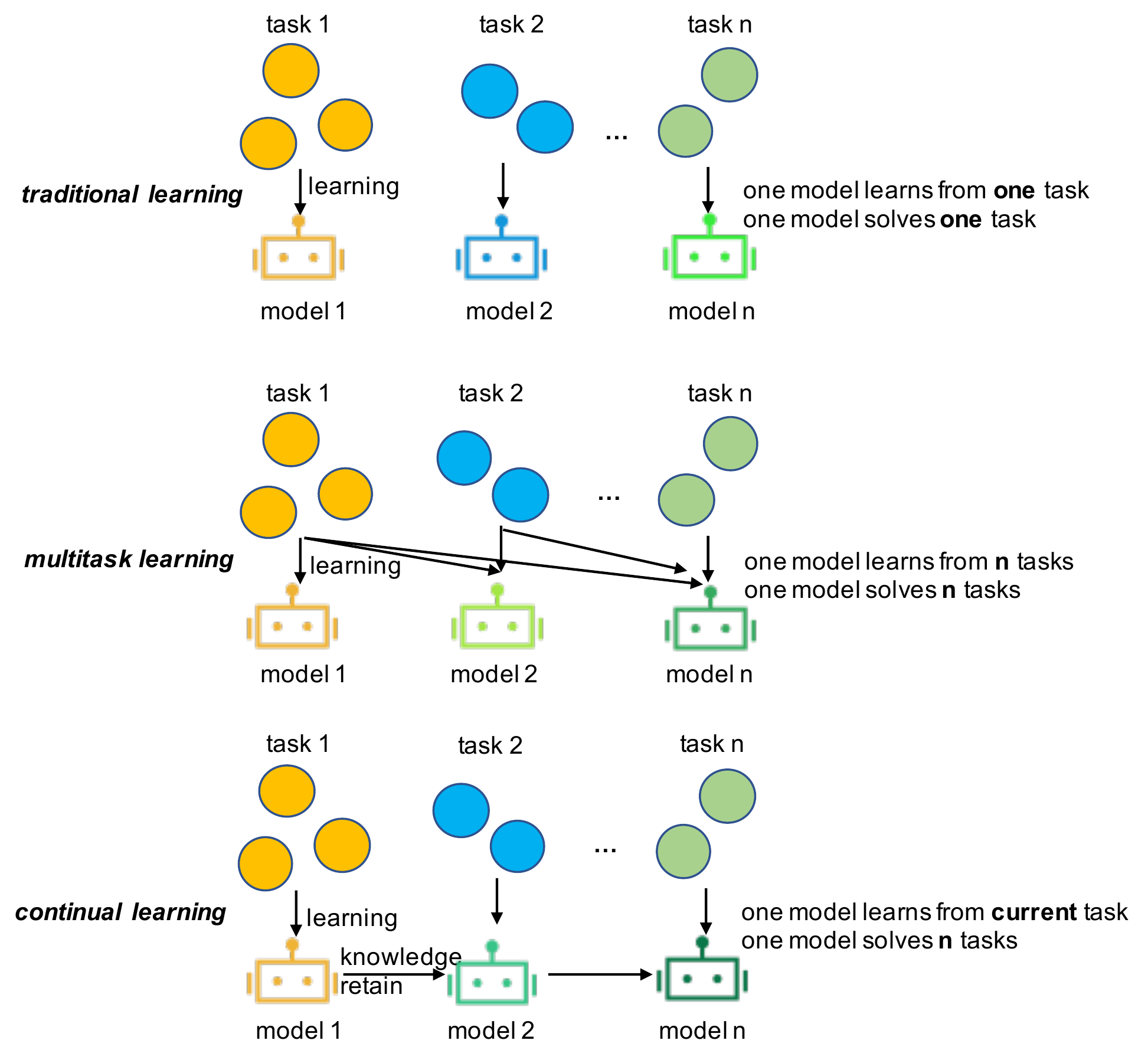}\caption{The difference among traditional learning, multitask learning, and continual learning paradigm.}
  \Description{Compare of different input}
  \label{fig_diff_paradigm}
\end{figure}

Figure~\ref{fig_diff_paradigm} illustrates the difference among traditional learning, multitask learning, and continual learning.
With traditional learning settings, a model is trained on a certain dataset and can only complete a single corresponding task.
Multitask Learning and Continual learning are similar in the sense that they both attempt to find a good solution across multiple tasks~\cite{mirzadeh2020linear}.
The main difference is that Multitask Learning has to keep access to all previous data. Once a new task and dataset are available, the model must be retrained on all historical datasets.
Therefore, the cost of Multitask Learning is much expensive than Continual Learning.
Formally, assuming that for task $t$, the dataset size is $d_{t}$ and the training cost (e.g. training time, computational resources) is $c_{t}$.
Then, in the task requirements growth progressively scenario, until task $t$ arrives, the total training cost for Multitask Leaning is $\sum_{i=1}^{t}\limits{(t-i+1)c_{i}}$, and Multitask Learning needs to store all historical data $\sum_{i=1}^{t}\limits{d_{i}}$.
Whereas for Continual Learning, the training cost is $\sum_{i=1}^{t}\limits{c_{i}}$ and it only needs to maintain current task's dataset.
Briefly, Multitask Learning is a good choice for the issues that tasks and data are both changeless, i.e., task requirements are clear, and all data are available at the beginning stage, rather than for our paper's focused problems.

The major challenge of Continual Learning is catastrophic forgetting (or catastrophic interference), which indicates that artificial neural networks tend to abruptly ``forget'' the knowledge of previously learned tasks~\cite{kirkpatrick2017overcoming}.
There are three main families of methods to mitigate this problem: Rehearsal, Regularization, and Architectural methods.
Rehearsal methods rely on collecting a part of typical historical data, so that they can replay them in the future training~\cite{robins1995catastrophic,rebuffi2017icarl,ramalho2019adaptive}.
Regularization methods apply some constraints to model's parameter updating, in order to strike a balance between stability and plasticity~\cite{kirkpatrick2017overcoming,su2020gradient}.
Architectural methods attempt to dynamically change model's modular, however, model parameters will dramatically increase when the number of tasks grows~\cite{mancini2018adding,wen2020batchensemble}.
In this work, we focus on the hybrid of rehearsal and regularization methods.

\subsection{Prompt for Pre-trained Model}
A prompt is a piece of tokens inserted in the input, so that the original task can be formulated as a language modeling task.
Prompt is used to fill the gap between pre-trained tasks and the down-stream task, facilitating finetuning process. 
Following Raffel et al. and Khashabi et al.~\cite{raffel2019exploring,khashabi2020unifiedqa}, we manually design a set of prefixes as prompt to concatenate each input component.

\begin{figure*}[!h]
  \centering
  \includegraphics[scale=0.54]{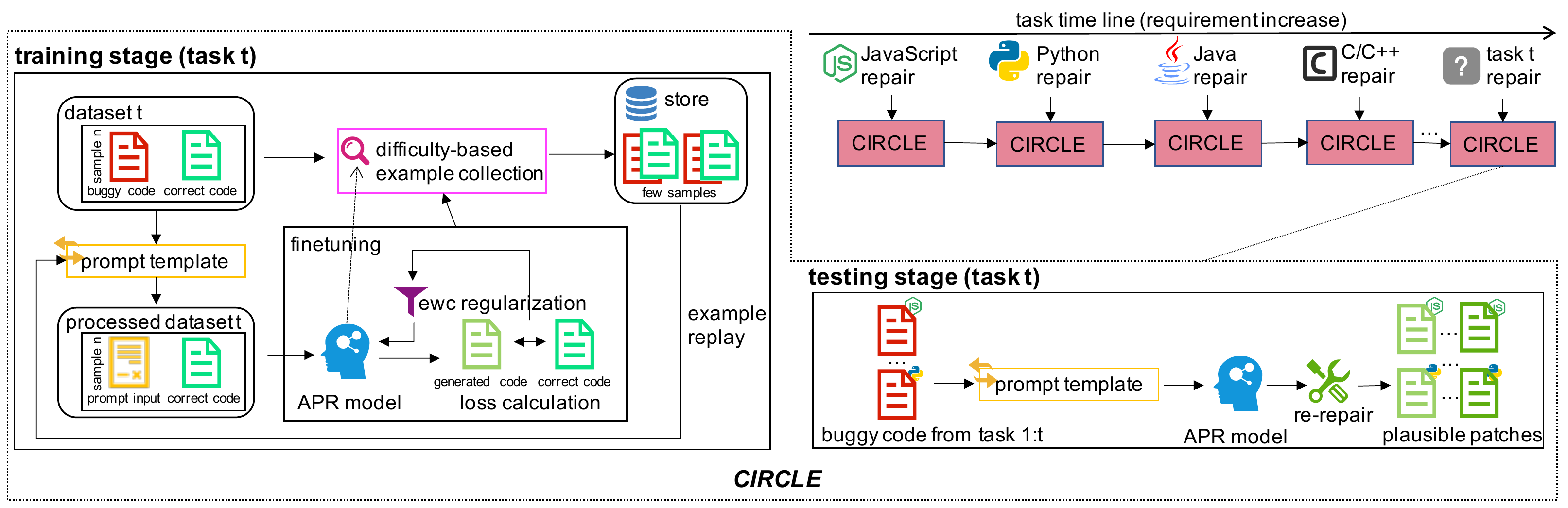}\caption{CIRCLE's overview.}
  \label{fig_architecture}
  \Description{CIRCLE architecture}
\end{figure*}
\section{Approach}
To address the limitations and challenges mentioned in Section~\ref{sec_intro}, we propose CIRCLE, a neural APR model that can continually learn defects fixing.
In this section, we introduce the design and implementation of CIRCLE.
First, we present the overview of CIRCLE in Section~\ref{sub_overview}.
CIRCLE aims to achieve both continual learning and multiple language repairing, which is more complex and also more practical than previous DL-based APRs.
It is composed of five parts.
First, CIRCLE leverages a large pre-trained model as its model skeleton to gain the strong learning ability (Section~\ref{sub_t5}), meanwhile, it employs a prompt function to effectively finetune the pre-trained model (Section~\ref{sub_prompt}).
Then, CIRCLE uses a novel difficulty-based rehearsal method (Section~\ref{sub_er}) and a parameter importance-based regularization (Section~\ref{sub_ewc}) to cope with forgetting problem.
Finally, a simple but effective re-repairing approach is utilized to erase generation errors caused by crossing languages (Section~\ref{sub_post}).

\subsection{Overview}\label{sub_overview}
Figure~\ref{fig_architecture} presents the overview of our approach.
As shown in the upper right part, CIRCLE can deal with task requirements increasing due to its continual program repair learning ability.
Without loss of generality, we assume new language program repair tasks arrive over time with their corresponding datasets.
CIRCLE automatically learns these tasks one by one based on task arriving order and does not need to retrain on or store the whole previous tasks' data.

For each task, CIRCLE consists of two stages: training stage and testing stage.
During training stage, a manually designed prompt function at first converts the repairing input into a fill-in-the-blank form.
Our training set is composed of two subsets: the current task corpus and a few examples selected from previous tasks.
Then, these processed data are fed into a T5-based APR model.
T5 utilizes a subword tokenization method to address out-of-vocabulary (OOV) problem.
Unlike previous work~\cite{jiang2021cure}, we keep the original tokenization vocabulary instead of building a new vocabulary using byte pair encoding (BPE)~\cite{sennrich2015neural} algorithm.
Because (1) we want APR model to inherit the natural language understanding ability and start learning repairing from a good initial point;
(2) BPE needs to count the subwords frequency, however, the frequency is dynamically changed crossing languages.
The T5-based APR generates candidate patches according to the prompted input and a loss function is applied to evaluate this generation.
To alleviate catastrophic forgetting, CIRCLE should carefully update its parameters.
Therefore, an EWC regularization is employed to compute the ``importance'' of each parameter for previous tasks, avoiding too much change of these more ``important'' parameters.
Finally, APR model is updated based on the loss and EWC regularization.
When the training is converged, we use this well-trained APR model to select a small set of examples from current task corpus via a novel data selection scheme.
These selected examples are stored to be replayed in the forthcoming task training.

During each task's inference stage, the APR model receives all seen languages' buggy codes and repairs them through generating a group of candidate patches.
In multiple programming language repairing scenarios, APR model is easy to incorrectly generate some keywords, since they have very similar semantic meanings.
For example, Java and JavaScript's ``null'' is similar to Python's ``None'' in both program and natural language aspects.
In some cases, our APR model will make mistakes about these keywords.
However, the number of such keywords is not too much.
We simply build a simple map to convert them after model's generation.

\subsection{Prompt based Data Representation}\label{sub_prompt}
\begin{figure}[!t]
  \centering
  \includegraphics[width=3.3in]{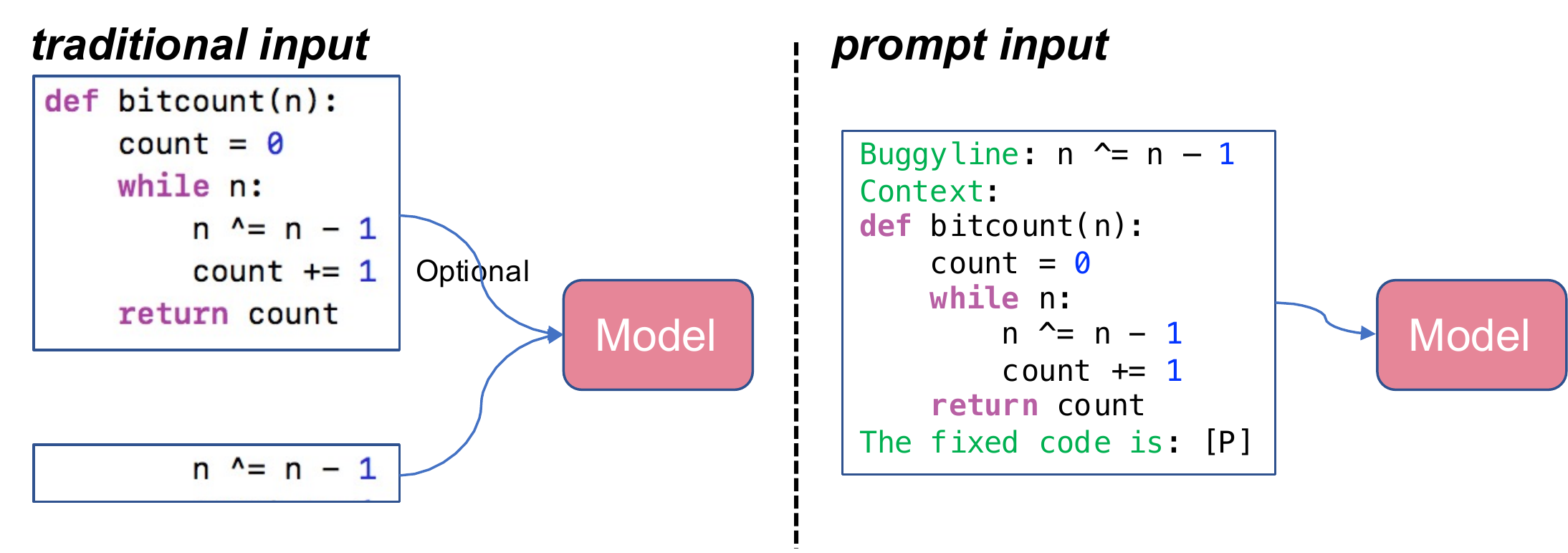}\caption{The comparison of traditional input and our prompt-based input. The \textcolor{green}{green} text represents prompt, which indicates the semantic meaning of each input component. The prompt input formulates bug-fixing task as fill-in-the-blank task, which is similar to T5's pre-trained task.}
  \Description{Compare of different input}
  \label{fig_inp_compare}
\end{figure}
The input of CIRCLE is composed of two parts: the buggy code and the surrounding context code.
Traditional works attempt to seperately encode these two parts and then merge the encoding vectors~\cite{lutellier2020coconut,jiang2021cure}.
However, how to effectively fuse these seperated encoding vectors and eliminate the semantic gaps between two encoders is still worth discussing.
Recently, Raffel et al.~\cite{raffel2019exploring} propose a text-in-text-out input format, which concatenates different input components with some prefixed prompt.
This mechanism is proved to be useful for finetuning pre-trained model in down-stream tasks~\cite{liu2021pre,keskar2019ctrl}.
In light of this, CIRCLE employs a manually designed prompt template to convert buggy code and corresponding context into a unified fill-in-the-blank format.
As illustrated in Figure~\ref{fig_inp_compare}, we utilize ``Buggy line:'', ``Context:'' denote the buggy line and context codes, and then we use ``The fixed code is:'' to guide pre-trained model generate fixed program according to the previous input.
Since T5 is pre-trained in fill-in-the-blank tasks with natural language, it is more natural for it to finetune on the prompted data.

In addition, CIRCLE utilizes subword tokenization method to address OOV problem.
But we do not newly build a token vocabulary because we want to fully exploit the pre-trained knowledge from T5 and the frequency of tokens from the whole datasets is not available as we mentioned in Section~\ref{sub_overview}.
In other words, finetuning T5 in APR task can be viewed as ``domain-adaptive'' task to some extent in this paper, i.e. the gap between down-stream task and pre-trained task is close.

\subsection{T5 as APR Model Skeleton} \label{sub_t5}
T5~\cite{raffel2019exploring} is a kind of encoder-decoder transformer~\cite{vaswani2017attention} model pre-trained in many tasks on over $750$ GB datasets, achieving state-of-the-art performance on a variety of NLP tasks.
Mastropaolo et al. ~\cite{mastropaolo2021studying} show that T5 can also perform well in many code-related tasks.
Although they also employ T5 to solve an APR problem, we are the first to use T5 for APR in the continual learning scenarios and to solve the APR task with multiple languages simultaneously.

The encoder of T5 is a stack of transformer blocks, each of which contains two subcomponents: a multi-head self-attention layer followed by a position-wise feed-forward network.
Layer normalization~\cite{ba2016layer}, residual connectors~\cite{he2016deep}, and dropout operation~\cite{srivastava2014dropout} are also applied between each subcomponent to stabilize the training process.

The decoder of T5 is much similar to the encoder but it has attention mechanism after each self-attention so that it can attend to the output of the encoder.
In addition, the attention is causality-enabled to avoid information leaking during decoding.

T5 comes in different sizes. 
In this paper, we do not modify the vocabulary size and use the pre-trained ``t5-base'' as the training starting point.

\subsection{Difficulty-based Example Replay} \label{sub_er}
One of the significant contributions of CIRCLE is that it enables APR models to continually learn bug fixing.
To achieve this, CIRCLE incorporates a difficulty-based example replay and an EWC-based regularization to avoid forgetting.
In this part, we introduce the novel difficulty-based example replay.

The core idea of example replay is to retain a small set of samples from previous datasets, and replay these samples in later training.
As a result, model can avoid severely forgetting meanwhile does not need to retrain on whole historical data.
Straightforwardly, the effectiveness of such rehearsal method largely relies on the selected examples.
In this work, we propose to select the representative and diverse data based on difficulty.
The intuition is that difficult data might be more informative and useful for improving current model.
Since model has poor performance during previous training with these difficult data, it might incline to first forget the pattern learned from those data.

The difficulty selection criterion is as follows:
\begin{equation}
  d_{t}(x_{i}^{t}, y_{i}^{t}) = \frac{L(x_{i}^{t}, y_{i}^{t}|\theta_{t})}{\left|y_{i}^{t}\right|} \label{eq_diff_cri}
\end{equation}
where $x_{i}^{t}$ and $y_{i}^{t}$ are the $i$-th data pair in task $t$.
$\theta_{t}$ is the model that already well-trained in task $t$.
$L(\cdot|\theta_{t})$ is the loss function (e.g. cross-entropy) conditioned on model parameter $\theta_{t}$.
Since long sentences tend to accumulate higher loss values, we use the inverse of $y$'s length as the normalization factor.
Basically, Eq.~\ref{eq_diff_cri} reflects the confidence of model $\theta_{t}$ when repairing the data $x_{i}^{t}$, where a higher value indicates $x_{i}^{t}$ is more challenging to learn.
We collect $N$ samples that can maximize Eq.~\ref{eq_diff_cri} from task $t$'s datasets as the difficult example set $E_{t}$.
$N$ is much smaller than the whole dataset $D_{t}$.
Consequently, for each task $i$, we maintain a corresponding example set $E_{i}$.

During continual training of APR, the previously selected difficult example set $E_{1:t}$ are integrated to the coming task $t+1$'s dataset.
Therefore, the objective of training is to find $\theta_{t+1}$ that can minimize the loss on the combined dataset:
\begin{equation}
  \theta_{t+1} = \mathop{argmin}\limits_{\theta_{t}} \sum\limits_{(x,y)\in D_{t+1} \cup E_{1:t}} L(x,y|\theta_{t})
\end{equation}

\subsection{Sampling-based EWC Regularization} \label{sub_ewc}
Since the size of selected example set $|E_{1:t}|$ should be as small as possible to reduce computation and storage costs, the effects of remembering old patterns are not strong enough.
Therefore, we further apply a constraint to avoid model losing previous knowledge.
This constraint is based on Elastic Weight Consolidation (EWC)~\cite{kirkpatrick2017overcoming}, which is widely adopted to overcome catastrophic forgetting in continual learning.
EWC imposes restrictions on parameters according to the importance of the parameters for previous tasks.
The more importance the parameters in previous tasks, the more tightly EWC restricts their updating.
EWC uses Fisher Matrix as a measure of weight importance, the calculation is as follows:
\begin{equation}
  F_{i} = \bigtriangledown^{2}L(x,y|\theta_{t-1,i}) \label{eq_fisher}
\end{equation}
$F_{i}$ measures the importance of $i$th parameter in $\theta_{t-1}$. 
$F_{i}$ is used to regularize model's updating:
\begin{equation}
  EWC(\theta_{t}) = \sum \limits_{i}\lambda F_{i}(\theta_{t,i}-\theta_{t-1,i})^{2}
\end{equation}
where $\lambda$ is hyper-parameter controls the contributions of EWC regularization.

However, directly calculate Eq.~\ref{eq_fisher} is unscalable, since it needs to access to all historical data.
To efficiently approaximate EWC, we only calculate Fisher Matrix on the data collected by our difficulty-based selection scheme, i.e. $(x,y)\in E_{1:t}$.
Furthermore, we only sample $M$ items from $E_{1:t}$ to further reduce the computation costs.

Finally, the objective function of training is changed to:
\begin{equation}
  \theta_{t+1} = \mathop{argmin} \limits_{\theta_{t}} (EWC(\theta_{t}) +\sum\limits_{(x,y)\in D_{t+1} \cup E_{1:t}} L(x,y|\theta_{t}))
\end{equation}

\subsection{Cross Language Re-repairing} \label{sub_post}
After continual training, CIRCLE learns the latent repairing patterns of multiple language and can generate correct patches.
However, unlike natural languages, programming language has strict formal grammar.
Therefore, CIRCLE still faces three problems due to the complication of crossing language repairing.
The first one is that CIRCLE has possibility to generate keywords that have the same semantic meaning but belongs to other languages.
We name this problem as ``keywords mismatch''.
For example, ``None'' in Python has the same semantic meaning to ``null'' in Java.
In some cases, CIRCLE incorrectly generates ``None'' when fixing Java bugs.
To address this mismatch problem, we build a simple mapping table to convert these mismatch words to corresponding one.
Table~\ref{tb_exp_keywords} presents some examples from our keywords mapping table.
\begin{table}[!ht]
  \centering
  \caption{Examples from keywords mapping table.}
  \begin{tabular}{lcccc}
  \toprule[1pt]
    & \textbf{C}      & \textbf{Java} & \textbf{JavaScript} & \textbf{Python} \\ \hline
  1 & NULL            & null          & null                & None            \\ \hline
  2 & max             & Math.max      & Math.max            & max             \\ \hline
  3 & min             & Math.min      & Math.min            & min             \\ \hline
  4* & -\textgreater{} & .             & .                   & .               \\ \hline
  ... &    ...          &    ...                &     ...  & ...         \\ 
  \bottomrule[1pt]
  \end{tabular}\label{tb_exp_keywords}
  \end{table}

The other problem is ``format mismatch'', which refers to generating correct tokens but with incorrect form.
The typical example is that CIRCLE tends to generate ``=='' as ``= ='' which will lead syntax error.
We simply build a regular expression to remove unnecessary blank space.

Finally, since T5 model's tokenizer is designed for natural language, although it applies wordpiece algorithms, it still slightly suffers out-of-vocabulary problem for a few rare symbols.
In our experiments, we find three symbols which are OOV tokens, however, they are frequently used in programming languages~\footnote{We test the following symbols: ``+ - * / \% ** // == != <> > < >= <= = += -= *= /= \%= //= **= \& ~ | \^{} << >> \{ \} \textbackslash{ } \textbackslash\textbackslash \@ \# \$ ( )'' only find three symbols meet this requirement: ``<'', ``\^{}'', and ``\{''.}.
When generating these symbols, T5 will simply generate unknown tokens, which is quite inappropriate.
The re-repairing mechanism replaces these unknown tokens with those special symbols.

\section{Experimental Setup}
In this section, we introduce the experimental design, including the research questions we studied, the training datasets, evaluation benchmarks, and implementation details in the experiments.

\subsection{Research Questions}
CIRCLE is designed to fix multiple language bugs and to achieve continually learning bug fixes. 
To this end, we explore the following research questions (RQ):
\begin{itemize}
  \item RQ1. Can CIRCLE effectively learn bug fixing in ``task requirements increase constantly'' scenario?
  \item RQ2. What is the performance of a single CIRCLE model compared to state-of-the-art APR methods?
  \item RQ3. What are the contributions of the different components of CIRCLE?
\end{itemize}

\subsection{Datasets}
Following previous works~\cite{lutellier2020coconut,jiang2021cure}, we directly use the CoCoNut's training data released on Github\footnote{https://github.com/lin-tan/CoCoNut-Artifact/releases} as our method's training corpus.
Same as these works, to make the evaluation realistic, we remove Java data committed after 2006.
The size of original datasets is very large: $3\,241\,966$, $480\,777$, $2\,735\,506$, and $3\,217\,093$ bug-fix pairs for Java, Python, C, and JavaScript, respectively.
Limitted of computation power, we randomly select a part of remaining data ($0.4$ million) for training.
The following experimental results indicate that training on such part of data still get great performance.
We utilize the prompt template to concatenate the context data and buggy data, meanwhile, we truncate the inputs whose length are longer than $512$ after subword tokenization.

For RQ1, to better align with continual learning scenario in real life, we assume that CIRCLE learns different languages repairing in the order of the language's popularity\footnote{the popularity is according to language's change rate on Github. \url{https://madnight.github.io/githut/\#/pull\_requests/2021/3}}: JavaScript $\rightarrow$ Python $\rightarrow$ Java $\rightarrow$ C.
This setting is reasonable since in practice, companies often attempt to build tools for the most popular part and then refine them for other parts.
The final trained CIRCLE model is used to compare with all state-of-the-art APR models to answer RQ2.

\subsection{Implementation Details}
All of our approaches are built based on PyTorch.
We use the HuggingFace~\cite{wolf2019huggingface} implementation version of T5 and utilize ``t5-base'' as the initial point, considering previous work recommendation~\cite{elnaggar2021codetrans,raffel2019exploring} and our devices' limits.
``t5-base'' model contains $12$ layers of transformer blocks and $12$ attention heads. 
The optimizer is AdamW~\cite{loshchilov2017decoupled} with $3e-4$ learning rate.
For each task (or splitted data in RQ2.), we train at most $20$ epochs, and if the validation loss does not decrease after $3$ epochs, the training process will be early stopped.
The batch size is $64$, the max length of input is set to $512$, and the $\lambda$ of EWC is $110000$.
The size of example set in difficulty-based example replay is restricted to $20000$, which is much smaller than the total size of training data.

In inference stage, we use beam search with $250$ beam size.
Meanwhile, we apply top-k and top-p sampling during each step's token selection.
Then, we re-repair the generated patches using the mapping table mentioned in Section~\ref{sub_post}. 
As a result, at most $1000$ candidate patches are created by CIRCLE.
For evaluation purpose only, following previous works~\cite{lutellier2020coconut,jiang2021cure}, three authors manually verify plausible patches (i.e. patches that successfully pass the test) based on ground truth patches (i.e., developer patches).
And the plausible patches is considered to be correct only if all three authors agree it is equivalent to ground truth data semantically.
All the training and evaluation of our methods are conducted on one CentOS 7.7 server with eight Tesla V100-SXM2 GPUs.

\subsection{Benchmarks and Baselines}~\label{sec:baseline}
We use five benchmarks with four popular programming languages: Defects4J~\cite{just2014defects4j} and QuixBugs~\cite{lin2017quixbugs} for Java, BugIDs~\cite{hanam2016discovering} for JavaScript, ManyBugs~\cite{le2015manybugs} for C, and QuixBugs~\cite{lin2017quixbugs} for Python,
all of which have been adopted in previous APR work~\cite{lutellier2020coconut, jiang2021cure, ghanbari2019prapr}.

In order to verify the continual learning ability of CIRCLE, i.e. RQ1, we train a T5 model in the traditional finetuning way as our baseline.
Finetuning way means that with the progress of tasks, model is initialized with checkpoint obtained from the last task and then, it is finetuned with current task's data.
We name this model as Finetuned-APR.
The training and inference parameters of Finetuned-APR is the same as CIRCLE.
The comparison between CIRCLE and Finetuned-APR indicates the superiority of our approaches in task streaming settings.

For RQ2, we employ five benchmarks commonly used for APR that contain realistic bugs. 
To enable sufficient evaluations, we compare {\toolname} against 30 APR techniques covering different programming languages and technique categories.
Specifically, all APR tools in the previous evaluation~\cite{lutellier2020coconut} and three recent state-of-the-art NMT-based tools ~\cite{jiang2021cure,Zhu2021recoder} are considered.

\section{Evaluation and Results}
In this section, we evaluate CIRCLE and answer to four research questions based on experimental results.

\subsection{RQ1: Can CIRCLE effectively learn bug fixing in ``task requirements increase constantly'' scenario?}
\begin{figure}[htbp]
  \centering
  \includegraphics[width=3.3in]{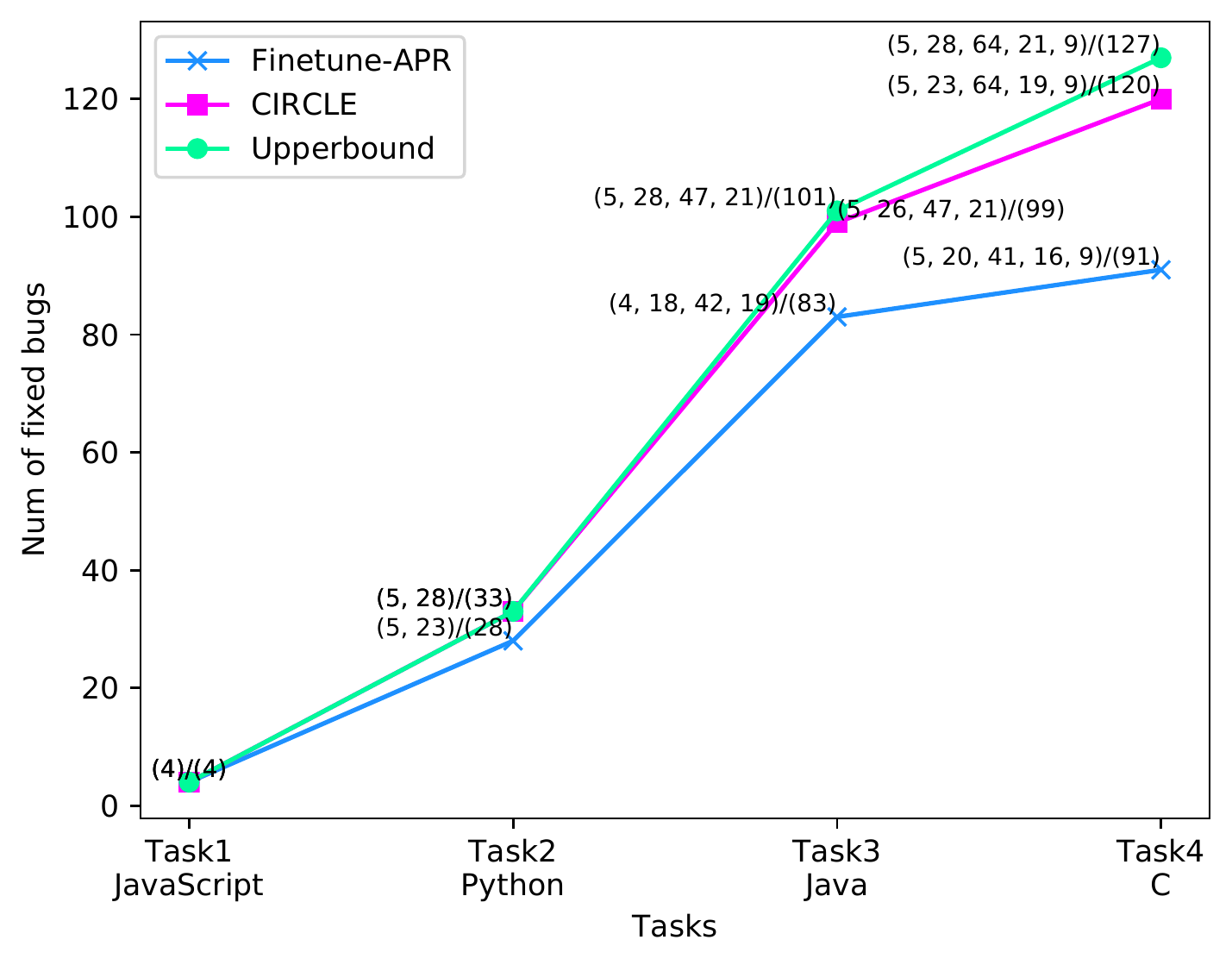}\caption{The performance trend comparison of traditional training approach and our CIRCLE in ``task requirements increase constantly'' scenario. The dot value (x)/(y) reports model's behavior on all seen tasks. The number in ``x'' shows the performance on each benchmark following the order: BugAID, QuixBugs-Py, Defects4J, QuixBugs-Java, and ManyBugs. ``y'' is the sum of ``x''.}
  \Description{Compare of different input}
  \label{fig_task_inc_comp}
\end{figure}

We compare CIRCLE with the Finetuned-APR which we mentioned in Section~\ref{sec:baseline}.
Note that the only difference between CIRCLE and Finetuned-APR is that the latter does not incorporate continual learning modules and is directly trained in a finetuning way.
Figure~\ref{fig_task_inc_comp} reports the performance trend of CIRCLE and Finetuned-APR with regarding to the task progress.
To be specific, during the learning progress, we account how many bugs the model can fix for both current task and previous learned tasks.
For example, after model learned Task 3 (i.e. Java) APR, we calculate the number of bugs it can fix on JavaScript (BugAID), Python (QuixBugs), and Java (Defects4J and QuixBugs) benchmarks.
In Figure~\ref{fig_task_inc_comp}, CIRCLE and Finetuned-APR have the same performance for the first task. 
Then, Finetuned-APR is gradually falling behind our CIRCLE.
Moreover, we can observe that: 
(1) For historical tasks, CIRCLE significantly outperform Finetuned-APR. For instance, after Task 3, Finetuned-APR can only repair $18$ bugs on QuixBugs-Py, in contrast, CIRCLE fixes $26$ bugs.
This observation supports the effectiveness of our proposed continual learning modules.
(2) Even for current tasks, CIRCLE still have better performance than Finetuned-APR. Take Task 2 as examples, Finetuned-APR correctly generates patches for $23$ bugs on QuixBugs-Py, however, CIRCLE generates correct codes for $28$ bugs.
This observation supports our argument that the underlying code understanding and patches construction abilities are largely common among programming languages.
As a result, if model does not severely forget previous obtained knowledge, this knowledge will help them in the following tasks.
To further demonstrate that the previous task's knowledge is helpful, we train a T5-based APR model for Python without learning any previous task.
It fixes $21$ bugs, which is worse than Finetune-APR's performance on Task 2.
This experimental result shows that although Finetune-APR tends to forget most previous knowledge, it still remembers a part of learned transferred knowledge and therefore, achieve better performance than the independently trained model.

Besides, we also draw the estimated upperbound in Figure~\ref{fig_task_inc_comp} to show the ``forgetting'' problem that our CIRCLE still suffered. 
At each task point, we choose the best performance that CIRCLE or Finetuned-APR achieved on each benchmark as the upperbound performance.
This upperbound curve indicates the degree of forgetting CIRCLE still suffers.
Figure~\ref{fig_task_inc_comp} shows that CIRCLE slightly falls behind the estimated upperbound compared to Finetuned-APR, further indicating the effectiveness of our continual learning strategies.

\subsection{RQ2: What is the performance of a single CIRCLE model compared to the dedicatedly trained state-of-the-art APR methods?}

\begin{table*}[htbp]

  \caption{Comparison with state-of-the-art techniques.
  Note that due to the the accessible of source code and time consuming of repair process, following the most of existing APR work \cite{ghanbari2019prapr,Zhu2021recoder}, we reuse the released results from the recent work \cite{lutellier2020coconut}. 
  Meanwhile, the results of most recent tools (i.e., CURE and Recoder) are extracted from the original papers.
  According to \cite{lutellier2020coconut}, the two duplicated bugs (i.e., Closure 63 and Closure 93) are excluded since they are same with Closure 62 and 92, respectively.
  (\dag) No plausible patch of Recoder is reported in the original work and five originally reported correct patches are false positives in the latest official repository (https://github.com/pkuzqh/Recoder).
  (\ddag) The number is calculated by the exact match of the generated patch and the developer patch, and indicates minimal number of correct patches.
  }
  \label{tab_rq3_performance}
  \begin{tabular}{ccclllll}
    \toprule[1pt]
  \multirow{3}{*}{FL} & \multirow{3}{*}{ID} & \multicolumn{1}{c}{\multirow{3}{*}{Tool}} & \multicolumn{2}{c}{Java} & \multicolumn{1}{c}{C} & Python & JavaScript \\
  \cmidrule(lr){4-5}\cmidrule(lr){6-6}\cmidrule(lr){7-7}\cmidrule(lr){8-8}
   & & \multicolumn{1}{c}{} & Defects4J & QuixBugs & ManyBugs & QuixBugs & BugAID \\
   & & \multicolumn{1}{c}{} & 393 bugs & 40 bugs & 69 bugs & 40 bugs & 12 bugs \\
    \hline
  \multirow{6}{*}{Standard} & T1 & Angelix & - & - & 18/39 & - & - \\
   & T2 & Prophet & - & - & 15/39 & - & - \\
   & T3 & SPR & - & - & 11/38 & - & - \\
   & T4 & Astor & - & 6/11 & - & - & - \\
   & T5 & LSRepair & 19/37 & - & - & - & - \\
   & T6 & DLFix & 29/65 & - & - & - & - \\
    \hline
  \multirow{8}{*}{Supplemented} & T7 & JAID & 9/31 & - & - & - & - \\
   & T8 & HD-Repair & 13/23 & - & - & - & - \\
   & T9 & SketchFix & 19/26 & - & - & - & - \\
   & T10 & ssFix & 20/60 & - & - & - & - \\
   & T11 & CapGen & 21/25 & - & - & - & - \\
   & T12 & ConFix & 22/92 & - & - & - & - \\
   & T13 & Elixir & 26/41 & - & - & - & - \\
   & T14 & Hercules & 49/72 & - & - & - & - \\
  \hline
  \multirow{17}{*}{Perfect} & T15 & SOSRepair & - & - & 16/23 & - & - \\
   & T16 & Nopol & 2/9 & 1/4 & - & - & - \\
   & T17 & (j)Kali & 2/8 & 1/2 & 3/27 & - & - \\
   & T18 & (j)GenProg & 6/16 & 0/2 & 2/18 & - & - \\
   & T19 & RSRepair & 10/24 & 2/4 & 2/10 & - & - \\
   & T20 & ARJA & 12/36 & - & - & - & - \\
   & T21 & SequenceR & 12/19 & - & - & - & - \\
   & T22 & ACS & 16/21 & - & - & - & - \\
   & T23 & SimFix & 27/50 & - & - & - & - \\
   & T24 & kPAR & 29/56 & - & - & - & - \\
   & T25 & AVATAR & 29/50 & - & - & - & - \\
   & T26 & FixMiner & 34/62 & - & - & - & - \\
   & T27 & TBar & 52/85 & - & - & - & - \\
   & T28 & CoCoNut & 44/85 & 13/ 20 & 7/- & 19/21 & 3/- \\
   & T29 & CURE & 57
  /104 & 26/35 & - & - & - \\
   & T30 & Recoder\dag & 64/- & - & - & - & - \\
\hline
   & T31 & CIRCLE & 64/182 & 19\ddag/- & 9\ddag/- & 23\ddag/- & 5\ddag/-\\
  \bottomrule[1pt]
  \end{tabular}
  \end{table*}
  
To evaluate the performance of {\toolname}, we compare it with state-of-the-art techniques, including the traditional and DL-based ones.
We adopt the final trained {\toolname} in RQ1 to perform repair tasks for five bug benchmarks across four programming languages.
Table \ref{tab_rq3_performance} shows the repair performance of a single {\toolname} model and the all selected baselines.
In Table \ref{tab_rq3_performance}, each cell is represented as $x/y$, where $x$ is the number of correct patches and $y$ is the number of produced plausible patches.
The results show that a single CIRCLE model achieves state-of-the-art performance in different languages.

As shown in Table \ref{tab_rq3_performance}, {\toolname}  fixes 120 bugs for all bug benchmarks in four programming languages. 
For Java benchmark, {\toolname} fixes 19 bugs on QuickBugs, outperforming CoCoNut and is competitive with CURE.
It is worthy noting that 19 bugs is calculated by exact match, which is usually the minimal of correct patch, while CoCoNut and CURE run all generated patches against the test suite.
Meanwhile, {\toolname} correctly repairs 64 bugs and outperforms all of the previous traditional APR techniques on Defects4J v1.2.
In particular, {\toolname} repairs 23.1\% (12 bugs) more bugs than the state-of-the-art traditional approach (e.g., TBar).
Meanwhile, {\toolname} can fix more bugs than most DL-based APR techniques (e.g., CoCoNut and CURE).
{\toolname} is also found to be competitive with the most recent DL-based approach, Recoder, which is reported to be the first DL-based APR approach that has outperformed the traditional APR approaches.
For other three language benchmarks, Table \ref{tab_rq3_performance} shows that {\toolname} is the best technique on two of the three benchmarks (i.e., fixs 19 bugs for QuickBugsPY, 9 bugs for ManyBugs, 5 bugs for BugAID, respectively), indicating that the repair ability across different programming languages with a single {\toolname} model.
It is worthy noting that most existing DL-based approaches (e.g., CURE) consider complex code-aware characteristics (e.g., code edits and abstract syntax tree)~\cite{Zhu2021recoder}, while {\toolname} treats the program repair process as a simple machine translation task on a sequence of tokens.
We only use around 4.13\% ($400~000 / 9~675~342$) of the data dataset compared with CoCoNut.
We also set the beam size as 250 while CURE’s beam is configured to 1000.
According to previous work\cite{tufano2019learning,niu2022spt}, a larger training set and beam size may lead to better repair performance.
Despite these points, {\toolname} still outperforms most of state-of-the-art APR techniques.
Thus, we highlight this direction of continual learning ability across multiple programming languages for automatic program repair.

\begin{figure}[t]

\centering

	\subfigure[Traditional approaches] {
		\includegraphics[width=0.47\columnwidth]{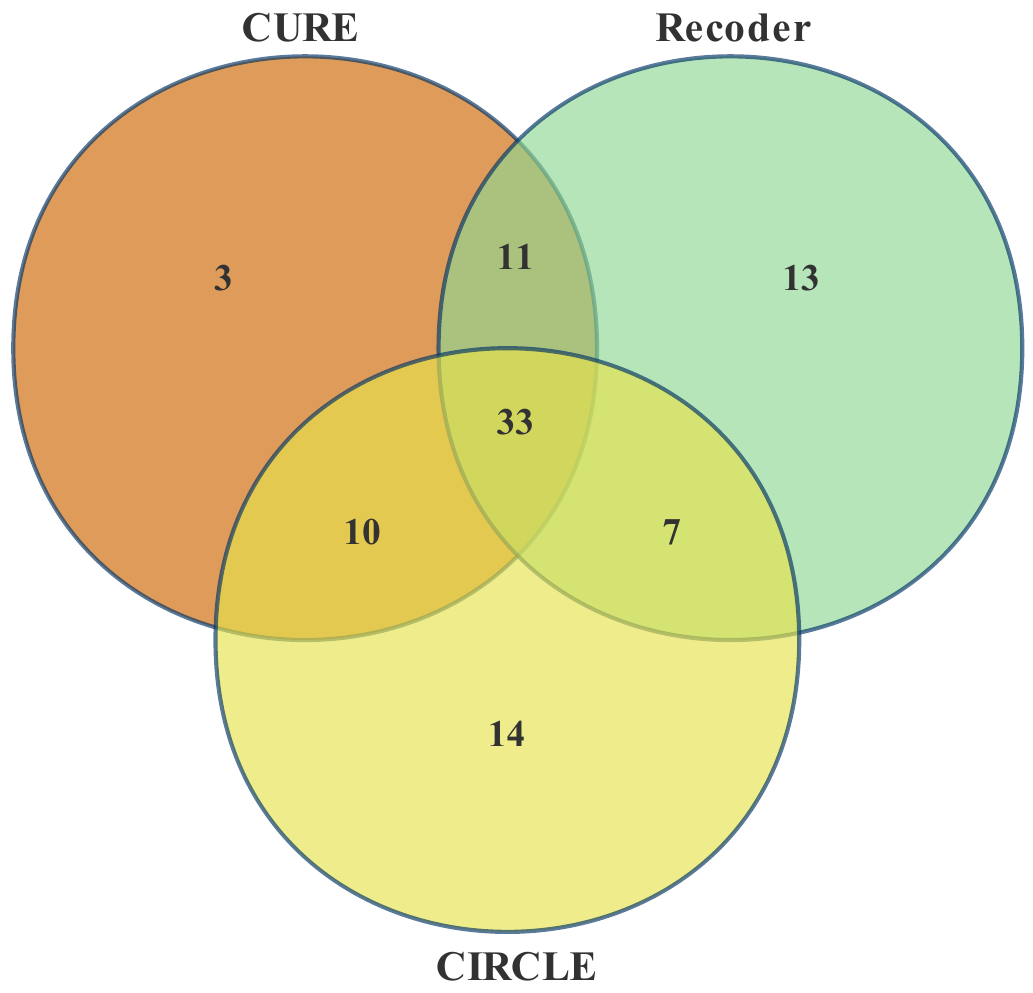}
	}
	\subfigure[DL-based approaches] {
		\includegraphics[width=0.47\columnwidth]{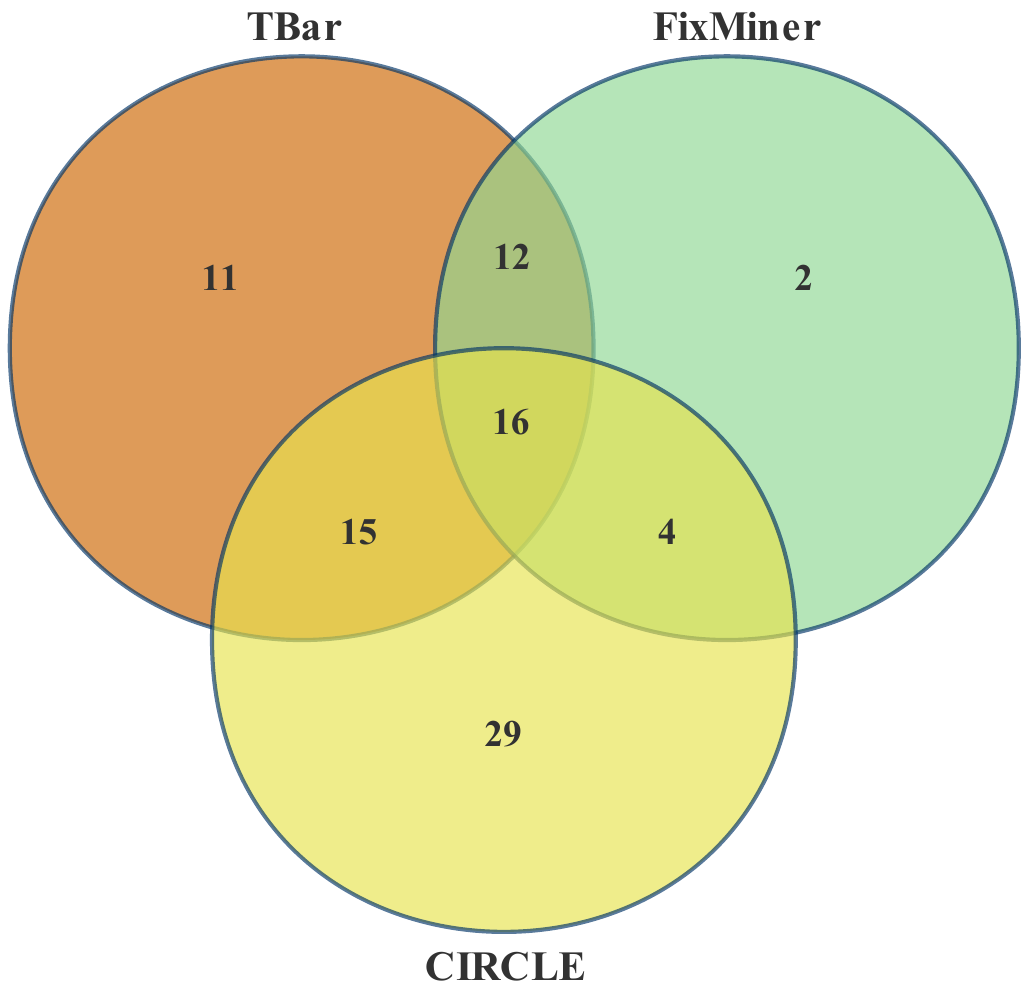}
	}
	\caption{The overlaps of the bugs fixed by different approaches}
	\label{fig:venn}
\end{figure}

To investigate what extent {\toolname} complements existing APR techniques, we further calculate the the overlaps of the bugs fixed by different techniques.
We focus on the Defecst4J benchmark as it is widely evaluated by most previous APR work \cite{li2020dlfix,Zhu2021recoder} and thus has rich performance results for overlap analysis.
Two best-performing traditional techniques (i.e., Tbar and FixMiner) and two best-performing DL-based techniques (i.e., Cure and Recoder) are selected.
As shown in Figure \ref{fig:venn}, {\toolname} fixes 29 and 14 unique bugs when compared with traditional and DL-based approaches. 
Moreover, {\toolname} fixes 33, 44, 21 and 24 unique bugs compared with TBar, FixMinder, CURE and Recoder, respectively. 
This result shows that {\toolname} is complementary to these best-performing existing techniques.

\subsection{RQ3: What are the contributions of the different components of CIRCLE?}
We investigate the impact of four components of CIRCLE:
(1) the prompt-based data representation;
(2) the cross language re-repairing;
(3) the lifelong module.

\subsubsection{Impact of prompt-based data representation}
According to recent research in NLP~\cite{liu2021pre}, adding prompt in input during finetuning can close the gap between the pre-trained task and down-stream task.
In this work, we use T5 as skeleton model, which is mainly pre-trained with natural language tasks and therefore is much different from the APR task.
To fill such gap, we concatenate context code and buggy code with some prompt words.
Intuitively, these simple prompt words mark the input with natural language, helping model better exploit its pre-trained knowledge.

\begin{figure}[!h]
  \centering
  \includegraphics[width=2.5in]{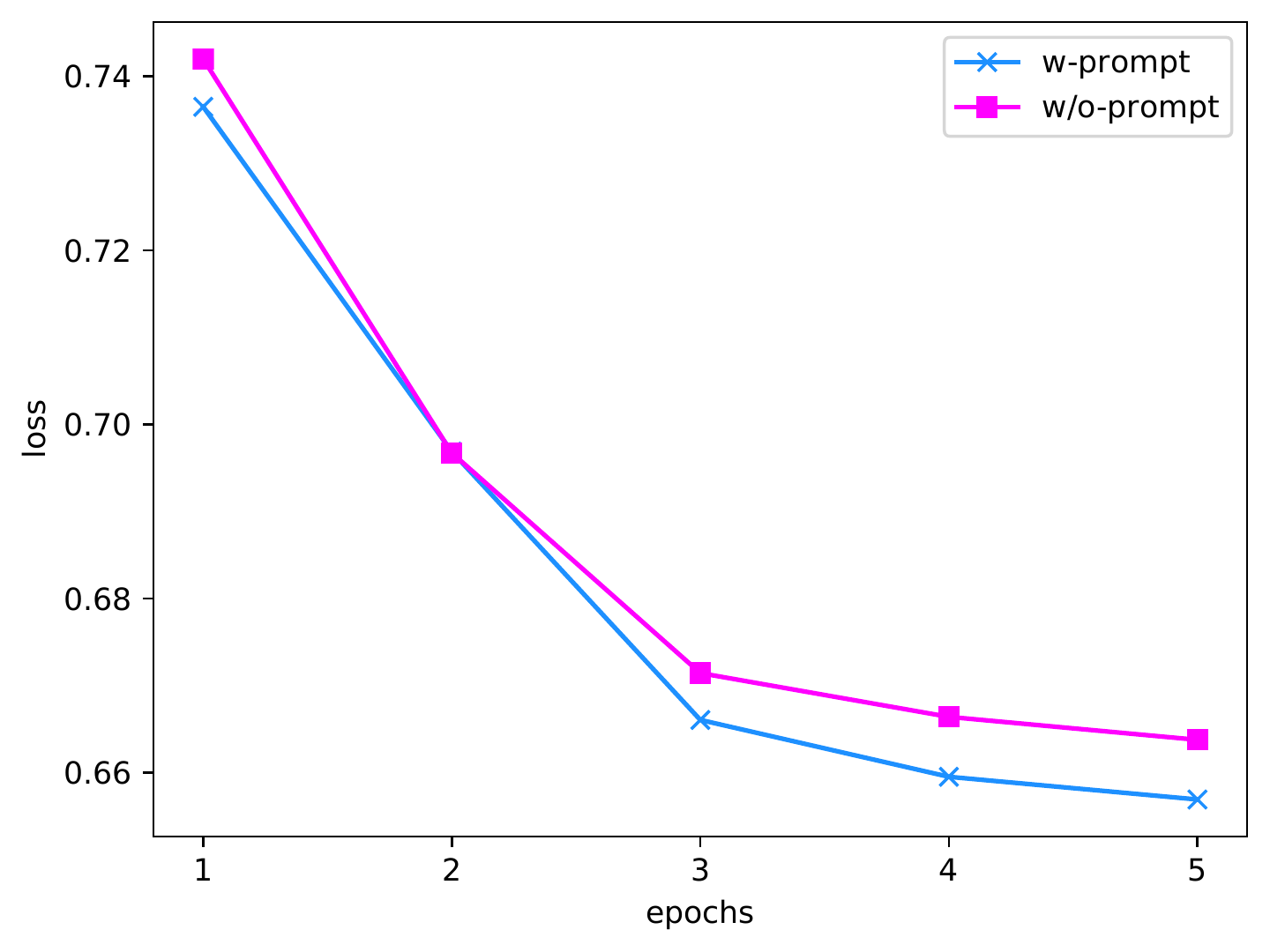}\caption{The validation loss of with/without prompt as input for the same model.}
  \Description{Compare of different input}
  \label{fig_wo_prompt}
\end{figure}

To investigate the actual impact of these prompt words, we train two CIRCLE model on the Java dataset.
One is fed with prompt-based input and the other is fed with no-prompt input.
Except the input form, all the parameters of these two models are the same.
Figure~\ref{fig_wo_prompt} presents the loss curve of these two models.
Within same number of epochs, the prompt-based inputs let model reach lower validation loss than no-prompt input.
In other words, the prompt-based input representation promotes the convergence of the model.

\subsubsection{Impact of re-repairing}

\begin{figure}[htbp]
\centering
\small
    \begin{tabular}{p{0.1cm}p{7.5cm}}

    \hline
    \multicolumn{2}{l}{\textbf{buggy line}} \\
    \textcolor{red}{-} &  while True: \\
    \hline
    \multicolumn{2}{l}{\textbf{patch generated by developer}} \\
    \textcolor{green}{+} &while queue: \\
    \hline
    
    \multicolumn{2}{l}{\textbf{patch generated before re-repairing}} \\
    \textcolor{green}{+} & while (NULL!= queue) \\
    \hline
    
    \multicolumn{2}{l}{\textbf{patch generated after re-repairing}} \\
    \textcolor{green}{+} & while (None!= queue) \\
    \hline
    
    \end{tabular}
\caption{Keywords mismatch example for BREADTH\_ FIRST\_ SEARCH from QuixBugs-Py}
\label{fig_no_rerepairing1}
\end{figure}

 \begin{figure}[htbp]
\centering
\small
    \begin{tabular}{p{0.1cm}p{7.5cm}}

    \hline
    
    \multicolumn{2}{l}{\textbf{buggy line}} \\
    \textcolor{red}{-} & if (typeof opt.default!='undefined') self.default(key, opt.default); \\
    \hline
    
    \multicolumn{2}{l}{\textbf{patch generated by developer}} \\
    \textcolor{green}{+} & if (typeof opt.default !== 'undefined') self.default(key, opt.default); \\
    \hline
    
    \multicolumn{2}{l}{\textbf{patch generated before re-repairing}} \\
    \textcolor{green}{+} & if (typeof opt.default!= = 'undefined') self.default(key, opt.default); \\
    \hline
    
    \multicolumn{2}{l}{\textbf{patch generated after re-repairing}} \\
    \textcolor{green}{+} & if (typeof opt.default!== 'undefined') self.default(key, opt.default); \\
    \hline
    
    \end{tabular}
\caption{Format mismatch example for INCORRECT\_ COMPARISON1\_2011 from BugAID}
\label{fig_no_rerepairing2}
\end{figure}

\begin{figure}[htbp]
\centering
\small
    \begin{tabular}{p{0.1cm}p{7.5cm}}

    \hline
    \multicolumn{2}{l}{\textbf{buggy line}} \\
    \textcolor{red}{-} & if (excerpt.equals(LINE) \&\& 0 <= charno \&\& charno < sourceExcerpt.length()) \{ \\
    \hline
    
    \multicolumn{2}{l}{\textbf{patch generated by developer}} \\
    \textcolor{green}{+} & if (excerpt.equals(LINE) \&\& 0 <= charno \&\& charno <= sourceExcerpt.length()) \{ \\
    \hline
    
    \multicolumn{2}{l}{\textbf{patch generated before re-repairing}} \\
    \textcolor{green}{+} & re-repairing: if (excerpt.equals(LINE) \&\& 0 <unk>= charno \&\& charno <unk>= sourceExcerpt.length()) <unk> \\
    \hline
    
    \multicolumn{2}{l}{\textbf{patch generated after re-repairing}} \\
    \textcolor{green}{+} & if (excerpt.equals(LINE) \&\& 0<= charno \&\& charno<= sourceExcerpt.length())\{ \\
    \hline
    
    \end{tabular}
\caption{Rare symbol example for Closure \#63 from Defects4J}
\label{fig_no_rerepairing3}
\end{figure}

As mentioned in Section~\ref{sub_post}, we employ re-repairing mechanism to solve ``keywords mismatch'', ``format mismatch'', and ``rare symbol'' problems.
In this subsection, we detailly analyze the influence of this re-repairing mechanism.

\begin{figure}[!h]
  \centering
  \includegraphics[width=3.in]{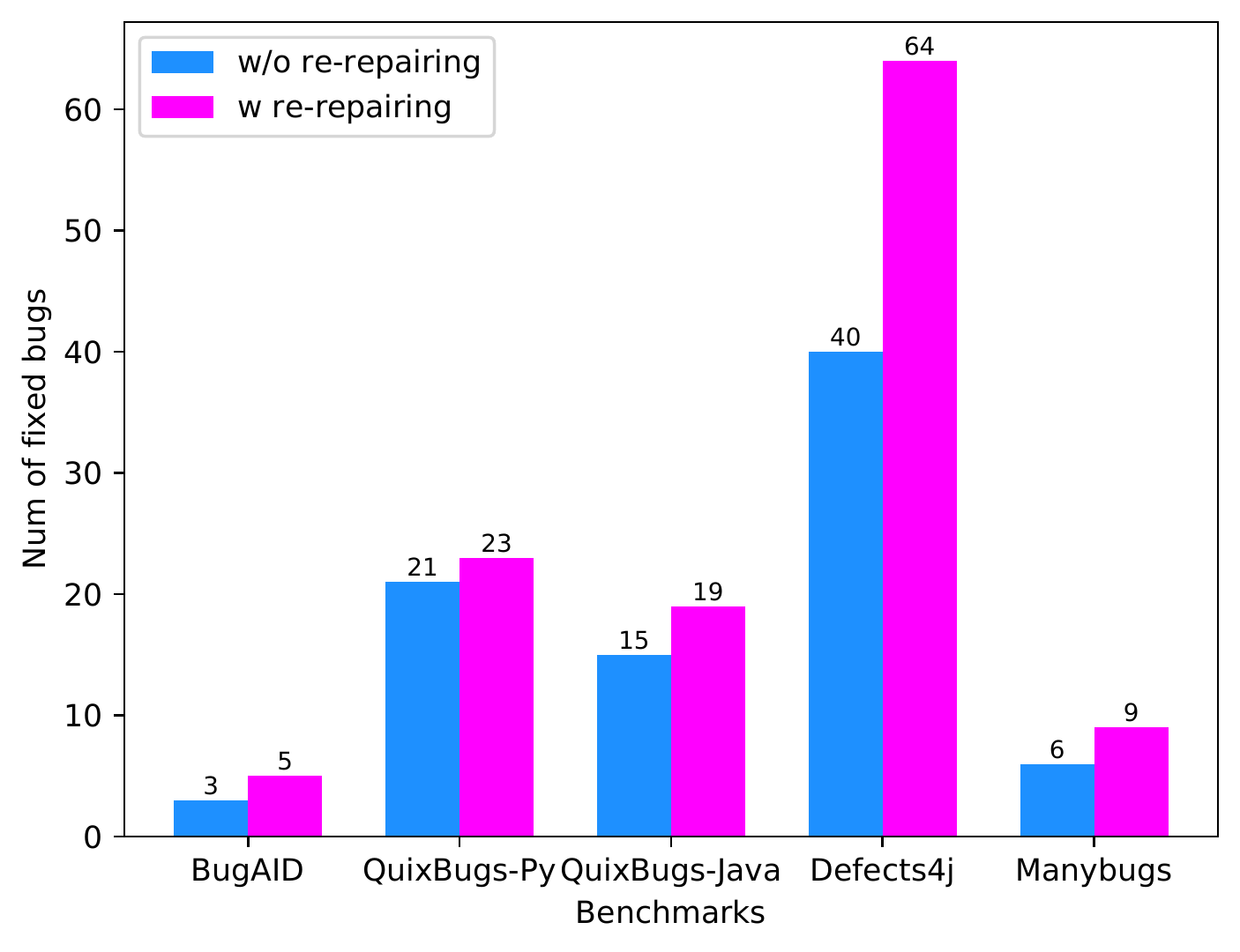}\caption{Number of fixed bugs on each benchmark with or without re-repairing mechanism.}
  \Description{Compare of different input}
  \label{fig_wo_rerepairing}
\end{figure}

Figure~\ref{fig_wo_rerepairing} displays the statistics of fixed bugs on all benchmarks when using or not using re-repairing.
As it shows, the simple re-repairing mechanism improves the number of fixed bugs on all benchmarks, demonstrating the efficacy of this module.
Figure \ref{fig_no_rerepairing1}, \ref{fig_no_rerepairing2}, \ref{fig_no_rerepairing3}. contains three examples illustrating how re-repairing mechanism addresses the ``keywords mismatch'', ``format mismatch'', and ``rare symbol'' problems.
For the first example, CIRCLE generates ``NULL'' in python patch, because the meaning of ``NULL'' is much similar to ``None'' in both programming language and natural language.
Our re-repairing module corrects this mistake, letting the generated patch be equivalent to the ground truth code.
The second example shows the procession that our re-repairing removes incorrect blank.
In the last sample, re-repairing fill rare symbols in the unknown token position to generate the fixed code.
From these example, we can observe that T5 generally generates the correct patches, however, it could make some naive mistakes.
The re-repairing mechanism is designed to fix these mistakes.

\subsubsection{Impact of the continual learning module}

The lifelong module in CIRCLE is mainly performed by the combination of the difficulty-based example replay and the sampling-based EWC regularization.
As the EWC's value is calculated on the selected sample set, we analyze them together.
In deed, the ``Finetune-APR'' in RQ1 is the version of CIRCLE removed lifelong learning module.
In RQ1, we show the impact of continual learning module through comparing the overall performance of CIRCLE and Finetune-APR.
Here, we give a detailed example from our experiments to show the continual learning's influence in a sample-level.
As shown in Figure \ref{fig_no_lifelong}, when the java training set is first given, both {\toolname} and Finetune-APR can generate the correct patch, which is the same as the developer patch.
However, Finetune-APR forgets to generate the condition ``endIndex < 0'' when the new C/C++ training set is in involved.

\begin{figure}[htbp]
\centering
\small
    \begin{tabular}{p{0.1cm}p{7.5cm}}

    \hline
    \multicolumn{2}{l}{\textbf{buggy line}} \\
    \textcolor{red}{-} & if (endIndex < 0) \{ \\
    \hline
    
    \multicolumn{2}{l}{\textbf{patch generated by developer}} \\
    \textcolor{green}{+} & if ((endIndex < 0)  || (endIndex < startIndex)) \{ \\
    \hline
    
    \multicolumn{2}{l}{\textbf{patch generated by {\toolname}}} \\
    \textcolor{green}{+} & if (endIndex <0 || startIndex >endIndex)\{ \\
    \hline
    
    \multicolumn{2}{l}{\textbf{patch generated by Finetune-APR}} \\
    \textcolor{green}{+} & if (endIndex <startIndex)\{ \\
    \hline
    
    \end{tabular}
\caption{Continual learning example for Chart \#9 from Defects4J}
\label{fig_no_lifelong}
\end{figure}
\subsection{Case Study}
For multiple languages repairing, one common concern is that ``can model distinguish different meanings of the same keywords among different languages?''.
For example, ``def'' is a keyword in Python, but it can be a variable name in other languages.
Will model be confused with such phenomenon?
To investigate this question, we conduct case study on two samples selected from QuixBugs-Java and QuixBugs-Py.
Specifically, we choose the Java and Python version of BITCOUNT program as an example.
We replace the variable name ``n'' with ``def'' in Java version, and replace ``n'' with ``public'' in Python version.
As shown in Figure~\ref{fig_key_case_study}, text with blue color represents the variable name replaced with other languages' keyword.
Our CIRCLE can correctly generate the patch without influenced by the ambiguious meaning of other language's keywords.
\begin{figure}[htbp]
  \centering
  \small
      \begin{tabular}{p{0.1cm}p{7.5cm}}
      \hline
      \multicolumn{2}{l}{\textbf{context line (Java)}} \\
       & public class BITCOUNT \{ public static int bitcount(int \textcolor{blue}{def}) \{ int count = 0; while (\textcolor{blue}{def} != 0) \{ \textcolor{blue}{def} = (\textcolor{blue}{def} \^{} (\textcolor{blue}{def} - 1)); count++; \} return count; \}\} \\

      \hline
      \multicolumn{2}{l}{\textbf{buggy line}} \\
      \textcolor{red}{-} & def = (def \^{} (def - 1));\\
      \hline
      
      \multicolumn{2}{l}{\textbf{patch generated by developer}} \\
      \textcolor{green}{+} & def = (def \& (def - 1));      \\
      \hline
      
      \multicolumn{2}{l}{\textbf{patch generated by {\toolname}}} \\
      \textcolor{green}{+} & def = (def \&(def - 1)); \\
      \hline
      
      \multicolumn{2}{l}{\textbf{context line (Python)}} \\
       & def bitcount(\textcolor{blue}{public}): count = 0 while \textcolor{blue}{public}: \textcolor{blue}{public} \^{}= \textcolor{blue}{public} - 1 count += 1 return count \\

      \hline
      \multicolumn{2}{l}{\textbf{buggy line}} \\
      \textcolor{red}{-} & public \^{}= public - 1      \\
      \hline
      
      \multicolumn{2}{l}{\textbf{patch generated by developer}} \\
      \textcolor{green}{+} & public \&= public - 1      \\
      \hline
      
      \multicolumn{2}{l}{\textbf{patch generated by {\toolname}}} \\
      \textcolor{green}{+} & public \&= public - 1 \\
      \hline
      
      \end{tabular}
  \caption{Case study for keyword meaning's ambiguity.}
  \label{fig_key_case_study}
  \end{figure}

\section{Related Work}
\subsection{APR}
Over the past decade, researchers have proposed a variety of techniques to generate patches based on different hypotheses \cite{gazzola2017automatic, monperrus2020living}.
Following recent work \cite{zhang2022program, benton2021evaluating, liu2020efficiency}, we categorize them into four main categories: heuristic-based~\cite{Le2011Genprog, Martinez2016Astor,Yuan2018Arja}, constraint-based~\cite{Durieux2016Dynamoth,Xuan2016Nopol, mechtaev2016angelix}, template-based \cite{Koyuncu2020Fixminer,Liu2019Avatar,Liu2019Tbar} and DL-based repair techniques~\cite{li2020dlfix,Zhu2021recoder,lutellier2020coconut}.

Recently, DL-based repair techniques, which attempt to fix bugs enhanced by machine learning techniques, is getting growing attentionRecently,  due to the large available open-source code.
Tufano et al. \cite{tufano2019empirical} extensively evaluate the ability of adopting neural machine translation techniques to generate patches from bug-fixes commits in the wild.
Li et al.~\cite{li2020dlfix} adopt a tree-based RNN encoder-decoder model (i.e., DLFix) to learn code contexts and transformations from previous bug fixes.
Lutellier et al. \cite{lutellier2020coconut} propose a new context-aware NMT architecture (i.e., CoCoNut) that represents the buggy source code and its surrounding context separately, to automatically fix bugs in multiple programming languages.
Jiang et al.~\cite{jiang2021cure} propose a novel approach (i.e., CURE) combines a program language model, a code-aware search strategy, and a subword tokenization technique.
The results demonstrate CURE can outperform all existing techniques on two popular benchmark (i.e., Defects4J and QuixBugs) when published.
Recently, Zhu et al.~\cite{Zhu2021recoder} use a syntax-guided edit decoder (i.e., Recoder) with provider/decider architecture to ensure accurate patch generation.
Compared to existing work, {\toolname} is the first work that aims to address the the generalizability issue of APR by repairing  multiple languages in a lifelong learning scenario.

\subsection{Continual Learning}
The general concept and technical categories of Continual Learning has been introduced in Section~\ref{sec:cl}.
Recently, to better fit real life applications, where tasks and data always change, Continual Learning has been widely used in many Natural Language Processing (NLP)~\cite{biesialska2020continual,nguyen2017retaining,yuan2022unified} and Computer Vision (CV) areas~\cite{parisi2019continual}, such as: Named Entity Recognition~\cite{monaikul2021continual}, Neural Machine Translation~\cite{cao2021continual}, Dialogue Systems~\cite{liu2021lifelong}, Question Anwering~\cite{penas2021continuous}, Text Classification~\cite{huang2021continual,chen2019exploiting}, Image Classification~\cite{van2019three,masana2020class,delange2021continual}, and so on.
However, none of existing works studied Automatic Program Repair (APR) with Continual Learning before.
This paper is the first one to explore continually learning APR tasks.

The idea of continual learning starts in 1990s~\cite{thrun1998lifelong}. 
However, achieving continual learning is challenging because of catastrophic forgetting~\cite{french1999catastrophic,kirkpatrick2017overcoming}.
Most of recent continual learning works focus on dealing with the forgetting problem.
Generally, these methods can be classified into three categories: Rehearsal, Regularization, and Architectural methods.
Rehearsal methods aim to replay some selected data in the forthcoming task.
iCaRL~\cite{rebuffi2017icarl} is one of the most well-known rehearsal method. 
It selects training data using Herding techniques~\cite{welling2009herding}.
Based on the replaying idea, some works build generators for previous task to create pseudo data for future learning~\cite{hu2018overcoming,hu2021distilling}.
Regularization approaches replies on additional loss term to consolid learned knowledge.
The classical regularization method is EWC~\cite{kirkpatrick2017overcoming}.
It restricts the update of ``important'' parameters.
Except EWC, other regularization methods such as GEM~\cite{lopez2017gradient}, MAS~\cite{aljundi2018memory}, IS~\cite{zenke2017continual} are also widely used to tackle catastrophic forgetting.
Architectural approaches prevent forgetting by applying modular changes to neural network models~\cite{rusu2016progressive,mancini2018adding,pfeiffer2020adapterhub,wen2020batchensemble}.
However, they will dynamically increase models' parameters when the number of tasks grows up.
In this paper, we combine rehearsal method with regularization to achieve lifelong learning in APR.

\section{Conclusion}
In this paper, we propose CIRCLE, an automatic program repairing framework that can continually learn to fix bugs crossing various programming languages.
Specifically, CIRCLE consists of five components: a prompt-based representation, a T5-based model, a difficulty-based example replay, an EWC-based regularization, and a re-repairing mechanism.
The T5-based model is the skeleton of APR model. 
The prompt-based representation converts program repairing to fill-in-the-blank task, filling the gap between T5's pre-trained task and APR task.
The difficulty-based replay and EWC-based regularization are two lifelong strategies, enabling CIRCLE to continually update its parameters according to the incremental task requirements.
Finally, a simple yet effective re-repairing method is applied to eliminate the form error caused by multiple languages repairing.
To the best of our knowledge, it is the first time to construct an APR model simultaneously addressing multiple programming languages based on continual learning approaches.
We conduct extensive experiments with $4$ programming languages on $5$ benchmarks to demonstrate the effectiveness of our CIRCLE.
Experimental results show that our CIRCLE (1) can continually learn bug fixing crossing languages; (2) achieves state-of-the-art performance on all benchmarks using a single model.

\section*{Acknowledgement} 
This work is partially supported by the National Key Research and Development Program of China (2021YFB1715600), National Natural Science Foundation of China (No. 62141215), Australian Research Council Future Fellowship (No. FT210100624), and Australian Discovery Project (No. DP190101985).


\bibliographystyle{ACM-Reference-Format}
\balance
\bibliography{sample-base}


\end{document}